\documentclass{desyproc}

\begin{document}
%------------------------------------
\title{Timelike Compton Scattering off the Proton: \\ 
 beam and/or target spin asymmetries.  
 }

\author{{\slshape Marie Bo\"er$^1$, Michel Guidal$^1$}\\[1ex]
$^1$Institut de Physique Nucl\'eaire, CNRS-IN2P3, Universit\'e Paris-Sud F-91406 Orsay, France 
}

 \contribID{256} 

% TO THE CONFERENCE EDITORS: 
% please update the follo wing information      
% before sending the template to the authors
\confID{8648}  % if the conference is on Indico uncomment this line
\desyproc{DESY-PROC-2014-04}
\acronym{PANIC14} % if you want the Acronym in the page footer uncomment this line
\doi  % if there is an online version we will register DOIs

\maketitle

\begin{abstract}
 We present a sample of results of our work to be published soon on  
 Timelike Compton scattering
  off the proton, in the framework
 of the Generalized Parton
 Distributions formalism. 
\end{abstract}

\section{Introduction}

\begin{wrapfigure}{r}{0.45\textwidth} 
\vspace*{-30pt}
  \begin{center}
    \includegraphics[width=0.4\textwidth]{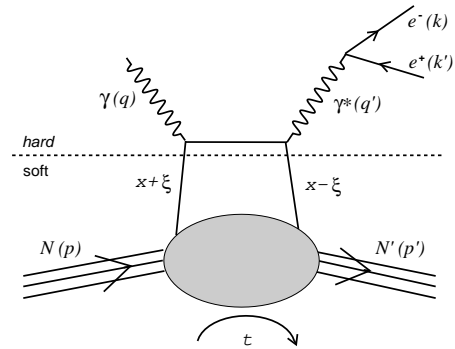} 
    \caption{Leading twist TCS diagram.  }
    \label{fig:TCS}
  \end{center}
  \vspace*{-15pt}
\end{wrapfigure}

More than 40 years after the discovery of pointlike components within the proton, 
 its quarks and gluons  structure   is still not well understood and is 
 still intensively studied.
 Hard exclusive processes on the proton 
 provide access to the Generalized Parton Distributions (GPDs)
  ~\cite{goeke,revdiehl,revrady,rpp}
  which 
 contain  informations about the   longitudinal momentum and the
spatial transverse distributions of partons inside the proton (in a frame where the nucleon has an ``infinite" momentum 
 along its longitudinal direction). 
 Such a hard exclusive process is the
  Deeply Virtual Compton scattering process  which corresponds to
  the reaction $\gamma^{(*)}P\to\gamma^{(*)}P$ and to 
the scattering of a high-energy virtual photon off a quark inside
 the proton.
There are two particular cases of deep Compton processes.
 ``Spacelike" 
 Deeply Virtual Compton Scattering  (DVCS) corresponds to the case  
  where the incoming photon 
   is emitted by a
    lepton beam  and 
    has a high spacelike virtuality and
       and where the final photon is real. 
   The DVCS process has been studied 
   for the past $\sim$15 years and  is   still  intensively studied  
   both theoretically and experimentally. The second particular
    case of deep Compton scattering is the 
 Timelike Compton Scattering (TCS) process. It  
corresponds to the case where 
the incoming photon is real and the final photon has a high timelike virtuality 
and decays into a lepton pair (see Fig. \ref{fig:TCS}). 
Contrary to DVCS, there is   no published experimental data yet for TCS. 
Both DVCS and TCS give access to the same proton GPDs in the QCD leading twist formalism.
  The study of TCS in parallel to DVCS  is a very powerful  way 
  to check the universality of GPDs and/or to study higher twist effects. 
 The reaction $\gamma P \to e^+e^- P$ 
also involves  the Bethe-Heitler process, where the incoming real 
photon creates a lepton pair, which then 
 interacts with the proton. It is not sensitive to the GPDs but to  the 
 form factors. It can   be calculated with a few percent accuracy.

\section{Amplitudes  and observables}

 The four vectors involved   are indicated in Fig.  \ref{fig:TCS}.
  According to QCD factorization theorems, at sufficiently large $Q'^2=(k+k')^2$
  (photon's virtuality), 
 we can decompose the TCS amplitude into a soft part, parameterized by the GPDs, and 
 a hard part, exactly calculable by Feynman diagrams techniques.
 We work in a frame   
  where the average protons and the average photons momenta, 
 respectively  $P$ and $\bar{q}$,  
 are collinear along the $z$-axis and    in opposite directions.
 We define the lightlike vectors along the positive and negative $z$ directions 
as $\tilde p^\mu = P^+/\sqrt{2} (1,0,0,1)$
 and $n^\mu = 1/P^+ \cdot 1/\sqrt{2} (1,0,0,-1)$, with $P^{+} \equiv (P^0 + P^3)/\sqrt{2}$. We have the properties  
  $\tilde{p}^2=n^2=0$ and $\tilde{p}\cdot n=1$. 
  In this frame, the  TCS amplitude  can be written 
   in the asymptotic limit (mass terms are neglected with respect to $Q'^2$)
    with the Ji convention for GPDs ~\cite{Ji97}: 
   \begin{eqnarray}
\label{eq:H_TCS}
&&T^{TCS} = -\frac{e^3}{q'^2} \:\bar{u}(k)\:
\gamma^{\nu} \:  \upsilon(k')\:
\epsilon^{\mu}(q)\:  \nonumber \\
&& \left[ \: \frac{1}{2}\:(-g_{\mu\nu})_{\perp}
\int\limits_{-1}^{1}dx\: \left(
\frac{1}{x-\xi-i\epsilon} + \frac{1}{x+\xi + i\epsilon}
\right)  
 .\left(
H(x,\xi,t) \bar{u}(p')\not{n}u(p)+ E(x,\xi,t)\bar{u}(p') i \sigma^{\alpha\beta}n_{\alpha} \frac{\Delta_{\beta}}{2M}\:u(p)
\right) \right.
\nonumber \\ 
&&\left. -\frac{i}{2}(\epsilon_{\nu\mu})_{\perp} 
\int\limits_{-1}^{1}dx\: \left(
\frac{1}{x-\xi-i\epsilon} - \frac{1}{x+\xi + i\epsilon}
\right) 
 .\left(
\tilde{H}(x,\xi,t)\bar{u}(p')\not{n}\gamma_5 \:u(p)
 + \tilde{E}(x,\xi,t)\bar{u}(p')\gamma_5 \frac{\Delta.n}{2M}\:u(p)
\right)\: \right],
\end{eqnarray}
 where $x$  is the quark longitudinal momentum fraction, $\Delta=( p'-p)$ 
 is the  momentum transfer, $t=\Delta^2$
  and 
  $\xi$ is defined as
  \begin{eqnarray}
\label{eq:Hxi} 
\xi= - \frac{(p-p').(q'+q)}{ (p+p').(q'+q)}. %\nonumber 
\end{eqnarray}    
 In Eq. \ref{eq:H_TCS}, 
 we   used the metric 
 \begin{eqnarray}
(-g_{\mu\nu})_{\perp}= -g_{\mu\nu} + \tilde{p}_{\mu}n_{\nu}   + \tilde{p}_{\nu} n_{\mu} \:\:, \quad
(\epsilon_{\nu\mu})_{\perp}=\epsilon_{\nu\mu\alpha\beta}\: n^{\alpha}\: \tilde{p}^{\beta}.
\end{eqnarray} 
 The Bethe-Heitler  
 amplitude reads: 
  \begin{eqnarray}
\hspace*{-0.1cm} T^{BH} = -\frac{e^3}{\Delta^2} \bar{u}(p')   
\left(
\gamma^{\nu} F_1(t)  + \frac{i \sigma^{\nu\rho} \Delta_{\rho} }{2\:M}F_2(t)
\right)
 u(p) \: \epsilon^{\mu}(q)  
\bar{u}(k)  
\left(
\gamma_{\mu}   
\frac{\not{k}-\not{q}}{(k-q)^2}
\gamma_{\nu}
+
\gamma_{\nu}   
\frac{\not{q}-\not{k'}}{(q-k')^2}
\gamma_{\mu}
\right)
\: \upsilon(k'), 
\end{eqnarray}
where $ F_1(t)$ and $F_2(t)$ are the proton Dirac and Pauli form factors. 
At fixed beam energy, 
the cross section of the photoproduction process depends 
 on four independant kinematic variables, which we
    choose as: 
  $Q'^2$,  $t$ and   the two angles $\theta$ and $\phi$ of the decay 
  electron in the $\gamma^*$ center of mass. The 
  4-differential 
  unpolarized cross section reads:
 \begin{eqnarray}
\frac{d^4\sigma}{dQ'^2dtd\Omega }({\gamma p \to p' e^+e^-})
= \frac{1}{(2\pi)^4}\frac{1}{64}\frac{1}{(2ME_\gamma)^2}\mid T^{BH} + T^{TCS}\mid^2,
\end{eqnarray}
where $\mid T^{BH} + T^{TCS}\mid^2$ is averaged over the target proton   
and beam polarizations and summed over the final proton spins. 
  \\
 
 We define the single and double spin asymmetries as:
 \begin{equation}
A_{\odot U} \: \: (A_{Ui})
= \frac{\sigma^{+} - \sigma^{-}}{\sigma^{+} + \sigma^{-}}, \quad
 A_{\odot i}= \frac{ (\sigma^{++} + \sigma^{--}) - 
( \sigma^{+-} +  \sigma^{-+}) }{\sigma^{++} + \sigma^{--}+
\sigma^{+-} +  \sigma^{-+}},
\end{equation}
where the first index of $A$ corresponds to the polarization state of the beam and the 
second one corresponds to the polarization state of the target. 
$A_{\odot U}$ is the circularly polarized beam spin asymmetry. The $+$ and $-$
 superscripts in $\sigma$ correspond to the two photon spin states, right and left polarized. 
 $A_{Ui}$ are the  single  target spin asymmetries
where the  $+$ and $-$ superscripts refer  to  the  
 target spin orientations  along the axis $i=x,y,z$. The 
axis $x$ and $y$   are perpendicular to the incoming proton direction (along the $z$-axis) in the $\gamma P$ center of mass frame and   
 are respectivelly  in the scattering plane  and perpendicular to this plane.
$ A_{\odot i}$ are the double spin asymmetries with a circularly   polarized beam   and with a polarized target. 
 We finally define the single  linearly  polarized beam spin asymmetry as  
\begin{equation}
A_{\ell U}(\Psi) = \frac{\sigma_{x}(\Psi) - \sigma_{y}(\Psi)}{\sigma_{x}(\Psi) 
+ \sigma_{y}(\Psi)},
\end{equation}
where 
  $\Psi $   is the angle  between the photon polarization vector and the
 $\gamma P \to \gamma^*P'$ plane and where 
$\sigma_{x}$ ($\sigma_{y}$)  indicate 
a photon   polarized in the $x$-($y$-)direction.

\section{Numerical results and sensitivity to GPDs }

\begin{figure}[hb]
\begin{center} 
\includegraphics[width=13cm,height=8.6cm]{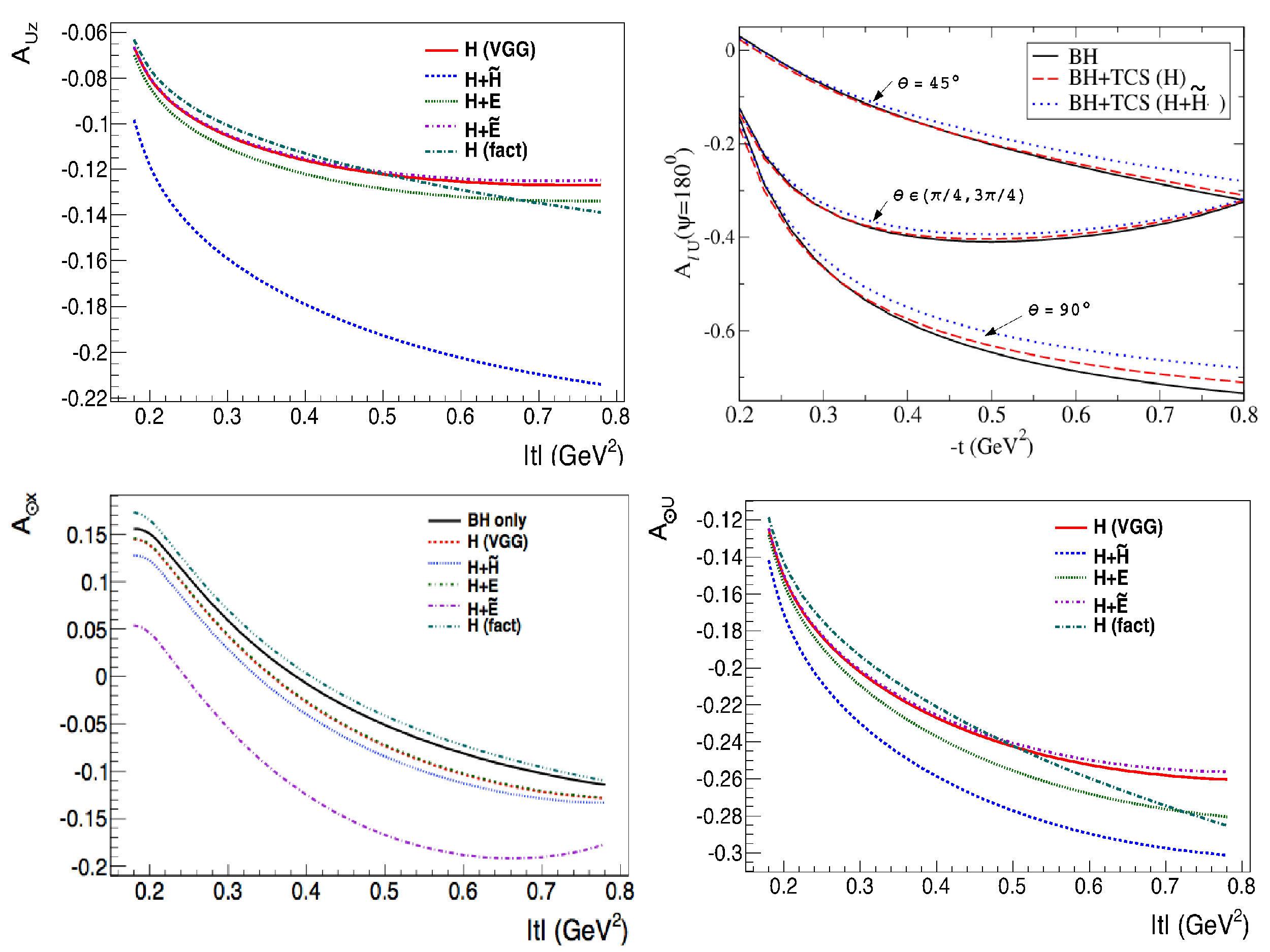}
\end{center}\vspace*{-0.2cm}
\caption{Spin asymmetries  as a function of $-t$.
 Top  left:  $A_{\odot U}$   for BH+TCS.    
 Top  right: $A_{\ell U}$   for BH and BH+TCS.
 Bottom left:  $A_{Uz}$  for BH+TCS.
Bottom right:   $A_{\odot x}$  for BH and BH+TCS.
 All   calculations are done at $\xi=0.2$,  $Q'^2=7$ GeV$^2$, $\phi=90^\circ$ and  $\theta$ integrated over $[45^\circ, 135^\circ]$.  $A_{\ell U}$ 
 is also shown for $\theta=45^\circ$ and $\theta=90^\circ$.
 }\label{fig:BSA}
\end{figure}

We performed our calculations  using the GPD parameterization of the VGG model
 ~\cite{Vanderhaeghen:1998uc,Vanderhaeghen:1999xj,Guidal:2004nd}.
    We focus here on the spin asymmetries. 
Figure \ref{fig:BSA}   shows the circularly (top row left) and linearly (top row right) 
polarized beam spin asymmetries as a function
 of $t$.
 One should note that $A_{\odot U} $ is particularly sensitive to the GPDs as 
 it is exactly $0$ for BH alone. It comes from the fact that this asymmetry is sensitive to the 
 imaginary part of the amplitudes and the BH amplitude is   purely real. We also show 
  $A_{\odot U} $  with a factorized-$t$ ansatz instead of a Reggeized-$t$ ansatz 
  for the $H$ GPD which illustrates the sensitivity to the GPD modeling.
   In contrast, the $A_{\ell U} $ asymmetry, which is strong,
    is dominated by the BH
   and the TCS 
   makes up only small deviations. Indeed, this 
   asymmetry is sensitive to the real part of the amplitudes.

We display in Fig. \ref{fig:BSA} (bottom row)
 two examples of asymmetries with a polarized target: 
   $A_{Uz}$    and 
  $A_{\odot x}$ (double spin asymmetry).
 We   present the results for TCS+BH with different GPDs contributions and 
 parameterizations.  
  All single target spin asymmetries are zero for the BH alone as they are proportional 
   to the imaginary part of the amplitudes. This  makes the  $A_{Ui}$
  asymmetries privileged observables to study GPDs. On the contrary, it is more
   difficult to  access GPDs with    double spin asymmetries as 
   the BH alone produces 
  a strong double spin asymmetry.

\section{Discussion}

 We  have presented a sample of our results to be published soon, namely
  the $t$-dependence of
  single and double spin asymmetries 
  for the 
  $\gamma P \to e^+e^- P$  reaction 
  which we analyzed in the framework of the GPD formalism.
  We didn't discuss this here due to lack of space but we 
  also   compared our unpolarized cross sections 
  and our single beam spin asymmetries with those 
  of the earlier work 
  of Refs~\cite{Berger:2001xd,Goritschnig} and they 
   are in agreement at the few percent level.
  We have introduced in our work the target polarization in order to define 
 the  single and double spin asymmetries with polarized targets. 
  We have
   also introduced some higher twist corrections and gauge invariance restoration terms. 
  \\
  
 As the BH contribution 
   alone doesn't contribute  to single target spin asymmetries and to 
   circularly polarized beam spin asymmetries, these  
   observables 
  are good candidates to study GPDs. Such measurements can be envisaged at the 
     JLab 12 GeV facility. In particular, 
   a proposal has been accepted for the CLAS12 experiment (JLab Hall B) to measure
   the unpolarized  BH+TCS cross section ~\cite{JLABprop}.
    The work that we presented 
   here  
     can open the way to new complementary experimental programs 
     with polarized beams and/or targets.

% ****************************************************************************
% BIBLIOGRAPHY AREA
% ****************************************************************************

\begin{footnotesize}
% IF YOU DO NOT USE BIBTEX, USE THE FOLLOWING SAMPLE SCHEME FOR THE REFERENCES
% ----------------------------------------------------------------------------

\end{footnotesize}

% ****************************************************************************
% END OF BIBLIOGRAPHY AREA
% ****************************************************************************

\end{document}